\documentclass[12pt,preprint]{aastex}

\tighten
\newcommand\be{\begin{equation}}
\newcommand\ee{\end{equation}}
\newcommand\jcd{Christensen-Dalsgaard}

\usepackage{epsfig}
\begin{document}
\shortauthors{Antia, Basu \& Chitre}
\shorttitle{Solar rotation and its gradient}

\title{Solar rotation rate and its gradients during cycle 23}
\author{H. M. Antia}
\affil{Tata Institute of Fundamental Research,
Homi Bhabha Road, Mumbai 400005, India}
\email{antia@tifr.res.in}
\author{Sarbani Basu}
\affil{Astronomy Department, Yale University, P. O. Box 208101,
New Haven CT 06520-8101, U.S.A.}
\email{basu@astro.yale.edu}
\and
\author{S. M. Chitre}
\affil{Centre for Basic Sciences, University of Mumbai, Mumbai 400098, India}
\email{kumarchitre@gmail.com}

\begin{abstract}
Available helioseismic data now span almost the entire  solar activity cycle 23
making it possible to study solar-cycle related changes of the
solar rotation rate in detail. In this paper we study how the solar
rotation rate, in particular, the zonal flows change with time.
In addition to the zonal flows that show a well known pattern
  in the solar convection zone, we also study changes in the radial and 
latitudinal gradients of the rotation rate, particularly
in the shear layer that is present in the immediate sub-surface layers of the Sun.
In the case of the zonal-flow pattern,
we find that the band indicating fast rotating region close
to the equator seems to have bifurcated around 2005.
Our investigation of the rotation-rate gradients show that the relative 
variation in the rotation-rate gradients 
is about 20\% or more
of their average values, which is much larger than the relative variation in
the rotation rate itself.
These results can be used to test predictions of various solar dynamo models.

\end{abstract}


\section{INTRODUCTION}
\label{sec:intro}

Helioseismology enables us to study dynamics of the solar interior
(e.g., Thompson et al.~1996; Schou et al.~1998, etc.). Early results have
shown that the differential rotation seen at the solar surface persists
till the base of the convection zone and becomes nearly uniform in the radiative
interior.
Also seen were two shear layers, one near the surface and the other near the
base of the convection zone. The latter, which is referred to as the tachocline
(Spiegel \& Zahn 1992), is generally believed to be the seat of the solar dynamo.
Superposed on the relatively smooth latitudinal variation are alternating
bands of slightly faster and slower rotation (Kosovichev \& Schou~1997).

Data for the
last 12 years covering most of the solar cycle 23 are now
 available from the Global Oscillation Network Group (GONG) project and
the Michelson Doppler Imager (MDI) instrument on board SOHO.  As a result, it is now possible
to study temporal variations of solar dynamics in detail.
Early works that used subsets of the data (e.g., Kosovichev \& Schou 1997;
Schou 1999; Howe et al.~2000a; Antia \& Basu 2000) 
had already established that there is significant variation in the rotation rate of the solar
interior.  The temporal variation
manifests as a system of bands with faster or slower than average rotation
rate at fixed depths in the upper layers of the  convection zone.
And particularly at low latitudes, the zonal flow bands appear to migrate towards the
equator with time, and appear to be similar in character to the torsional oscillations
discovered in the solar surface rotation rate (Howard \& LaBonte 1980;
Ulrich et al.~1988; Snodgrass 1992).
A subsequent work (Antia \& Basu 2001) found
that in addition to this pattern at low latitudes, there is another system
of zonal-flow bands at high latitudes ($> 45^\circ$) that migrate towards
the poles as the solar cycle progresses.
 With accumulation of more seismic data it has now become
clear that these zonal flows penetrate through most of the convection zone
(Vorontsov et al.~2002; Basu \& Antia 2003, 2006; Howe et al.~2006a).
Since the errors in the results obtained by 
inversions increase with depth, it is not yet clear whether there is any significant
temporal variation in the tachocline region near the base of the convection
zone ---  a region where the solar dynamo is believed to be operating.
Howe et al.~(2000b, 2007) reported  a 1.3 year oscillation in the equatorial region
around $r=0.72R_\odot$ during the rising phase of the solar cycle, but
that has not been confirmed by other studies (e.g., Antia \& Basu 2001;
Basu \& Antia 2003).

The origin of torsional oscillations or zonal flows is not fully
understood, but since
the zonal flow banded structure and the magnetic activity pattern
seem to be closely related, 
it is generally believed that zonal flows arise from a nonlinear
interaction between magnetic fields and differential rotation.
Thus the properties of solar zonal flows 
 should provide a constraint on theories of the solar dynamo.
Most dynamo models are kinematic, where the velocity field is
specified;  such models do not yield any information about temporal variations
in the velocity field. However, there are some non-kinematic dynamo models that  include the
Lorentz-force feedback on differential rotation, and these have been used to explain 
the zonal flow patterns.
Covas et al.~(2000, 2004), for example, considered an axisymmetric mean-field dynamo model
to study temporal variations of the rotation rate and of magnetic fields in
the solar interior. 
They found
temporal variations in the rotation rate that are qualitatively similar
to the zonal-flow pattern inferred from the seismic data.
Spruit (2003)  suggested that thermal fluctuations due to enhanced
emission of radiation by small-scale magnetic fields
can drive the zonal-flow pattern. His model produces zonal flows
with amplitudes that decrease  with increasing depth. Lanza (2007)
attempted to find the source of the zonal-flow pattern by using a model for
 angular momentum transport in the convection zone and by comparing
the results with observed patterns. His preliminary results led to
constraints on the amplitude, 
phase and location of the perturbations responsible for the observed torsional oscillations.
Rempel (2007) found that the poleward
moving branch of zonal flows at high latitude may be explained as a
response of the coupled differential rotation/meridional flow system to
periodic forcing in mid-latitudes by either Lorentz force or thermal
perturbations, with thermal fluctuations very likely playing a role in
the low-latitude equator-ward moving branch of the zonal flow.

Most seismic studies hitherto have investigated
temporal variations in the rotation rate itself. For the solar dynamo
models, however,
spatial gradients of this angular velocity are more important.
Although the radial gradient of the rotation rate in the outer shear layer
has been studied by 
Schou et al.~(1998), Basu \& Antia (2001), Corbard \& Thompson (2002), etc., 
a detailed study of the temporal variations of both the radial and
latitudinal gradients has not been undertaken so far.
This motivated us to further investigate changes in both the radial
and the latitudinal gradients of the solar rotation rate with a view to
studying temporal variations of the shear pattern.
We also study the time-derivative of the rotation velocity, which
should relate to the driving forces that are possibly responsible for the zonal flow
pattern.

The rest of this paper is organized as follows: we summarize the data and the
analysis technique in \S~\ref{sec:data}; our results are described in
\S~\ref{sec:results}, and the discussion of our results along with conclusions
are given in \S~\ref{sec:disc}.

\section{Data and technique}
\label{sec:data}

We use data obtained by the GONG (Hill et al.~1996) and MDI
(Schou 1999) projects in the present work.
These data sets consist of the mean frequency and the splitting coefficients
for different $(n,\ell)$ multiplets.
Only the odd-order splitting coefficients are required for determining
the rotation rate in the solar interior (e.g., Ritzwoller \& Lavely 1991).
These splitting coefficients are sensitive only to the north-south symmetric
component of the rotation rate and hence, that is the only component that can
be determined with these data. Consequently, all results shown in this work are
symmetric about the equator. There could
be an antisymmetric component of rotation present in the Sun, but that cannot be analyzed using
the global-mode data used in this work.
We use 120 data sets from GONG, each set
covering a period of 108 days. The first set starts on
1995, May 7 and the last set ends on 2007, May 15, with a
spacing of 36 days between consecutive data sets.
Thus there is a considerable overlap between neighboring data sets.
The GONG data sets include only p-modes up to $\ell=150$ and we use all modes
with frequencies between 1 and 3.5 mHz.
The MDI data sets consist of 56 non-overlapping
sets each obtained from observations taken over a   period of
72 days. The first set begins on 1996, May 1 and the last
set ends on  2007, October 6.
Contact with SOHO satellite was lost during June 1998
and the satellite was finally recovered during February 1999. During this
period there is only one data set and that covered a period of 60 days.
The MDI data sets include f-modes up to $\ell=300$ and p-modes up to
$\ell\approx200$. We use all modes
with frequencies between 1 and 3 mHz. Since the higher frequency modes often
cause problems, they have not been included in the analysis. The higher degree modes from
MDI are also known to give misleading temporal variations in some cases
(e.g., Antia et al.~2003). This is indeed found to be the case for
the radial gradient of the rotation rate and is discussed later.

We use a two dimensional Regularized Least
Squares (2D RLS) inversion technique as described by
Antia et al.~(1998) to infer the rotation rate in
the solar interior from each of the available data sets.
The first 8 odd-order splitting coefficients, $a_1,a_3,\ldots,a_{15}$, were
used in the inversions. 
In order to study the time-variation of the rotation rate we first
determine  the Sun's internal rotation rate using each data set,
and then take the time average of the results obtained using all data sets at each latitude
and depth. Since we treat the GONG and MDI sets separately, we  obtain
one average result using GONG sets and another using MDI sets.
To obtain the time-varying component of rotation, we subtract the  mean from the
rotation rate at any given
epoch. Thus,
\begin{equation}
\delta\Omega(r,\theta,t)=\Omega(r,\theta,t)-
\langle \Omega(r,\theta,t)\rangle,
\label{eq:zonal}
\end{equation}
where $\Omega(r,\theta,t)$ is the rotation rate as a function of
radial distance $r$, latitude $\theta$ and time $t$.
Here the angular
brackets denote the average over the time duration for which data are available.
As is the case for the time-averaged rotation,  the residual rotation rate is 
calculated separately for GONG and MDI data sets.

For most parts of this investigation we use the residual rotation velocity, $\delta v_\phi=
\delta\Omega r\cos\theta$, rather than the rotation rate $\delta\Omega$.
This time varying component, $\delta v_\phi$, is generally called the zonal flow.
The definition of zonal flows is somewhat ambiguous: some authors (e.g., Howard \&
LaBonte 1980)
have defined the zonal flows by subtracting the temporal average of
a smooth component of the rotation rate. The smooth component is generally
defined using three terms (constant, $\sin^2\theta$ and $\sin^4\theta$)
to calculate the latitudinal variation of rotation rate.
There is some difference in the resulting pattern in the two cases,
as has been described by Antia \& Basu
(2000). In this work we adopt the definition given by Eq.~(\ref{eq:zonal})
where the temporal mean of the full rotation rate is subtracted to obtain
the residuals. This definition has been used previously by other
investigators too (e.g., Howe et al.~2000a, 2005, 2006a; Antia \& Basu 2001;
Basu \& Antia 2003, 2006, etc.).

For the purpose of investigating the spatial gradients of the rotation rate, we use
$\Omega$ rather than $v_\phi$ and calculate the radial gradient,
($\Omega_{,r}\equiv\partial \Omega/\partial r$), as well as the latitudinal
gradient 
($\Omega_{,\theta}\equiv(1/r)\partial \Omega/\partial|\theta|$).
These derivatives are computed numerically.
Since the inferred
zonal flow pattern is symmetric about the equator, the radial gradient
is also symmetric. The latitudinal gradient on the other hand is antisymmetric
about the equator, and  hence it is 0 at the equator. To restore the
symmetry of the latitudinal gradient we, have defined the gradient with respect
to the absolute value of the latitude. With this definition, both $\Omega_{,r}$ and $\Omega_{,\theta}$
are symmetric about the equator.
Since the 2D~RLS technique used for inversion gives results that are
smooth in $(r,\theta)$, there is no difficulty in computing these derivatives numerically.
The errors in these derivatives are, of course,
larger than those in $\Omega$ itself.
The time-varying component of these derivatives is calculated by
differentiating the residual in rotation rate defined by Eq.~(\ref{eq:zonal}),
i.e.,$\delta\Omega_{,r}\equiv\partial \delta\Omega/\partial r$ and
$\delta\Omega_{,\theta}\equiv(1/r)\partial \delta\Omega/\partial|\theta|$.

We also calculate the derivative of the rotation velocity $v_\phi$ with respect to time
(note that $\partial v_\phi /\partial t=\partial\delta v_\phi/\partial t$),
since it is likely to give an indication of the forcing term that drives
the zonal-flow pattern.
For calculating time-derivatives numerically, we need $v_\phi$
at different times, which we obtain from independent data sets. It should be noted that
since the magnitude of the zonal-flow velocity is small, the
temporal derivative has significant errors. To get more reliable
estimates, we smooth $\delta v_\phi$ at each latitude and radius by assuming that
the temporal variation is periodic with a period of 11 years.
Since the variation is not strictly sinusoidal, we also include a few higher
harmonics and fit the following functional form to  $\delta v_\phi$:
\be
\delta v_\phi(r,\theta,t)=\sum_{k=1}^{k_{\rm max}} a_k(r,\theta)\sin(k\omega_0 t)
+b_k(r,\theta)\cos(k\omega_0 t)=\sum_{k=1}^{k_{\rm max}} A_k\sin(k\omega_0 t+
\phi_k)\;.
\label{eq:harm}
\ee
In the above equation, $\omega_0$ is the basic frequency for the fundamental component,
which is assumed to have a period of 11 years.
The coefficients $a_k$ and $b_k$ are determined by a least squares
fit at  fixed values of $(r,\theta)$. These coefficients determine the
amplitude, $A_k$ and phase, $\phi_k$ of the variation at any
point defined by the values of $(r,\theta)$ for each harmonic
of the 11 year period. We choose time $t=0$ to be the beginning
of GONG data sets and the phase is measured with respect to that epoch.
This zero-point is about a year before the minimum phase of solar activity.
Three terms are found to be sufficient to fit the
data. Similar fits were also used by Vorontsov et al.~(2002), Basu \&
Antia (2003, 2006) and Howe et al.~(2005, 2006a).
These fits are then used to calculate the time-derivative. These fits also enable us 
to extend the pattern beyond the epochs for which observations exist.
These extended patterns are often useful in comparing the observed pattern with those obtained
on the basis of dynamo models since dynamo-model predictions are traditionally shown over a period of 22 years.

\section{RESULTS}
\label{sec:results}

The rotation rate in the solar convection zone exhibits temporal variations. 
Near the solar surface the
pattern is observed to be similar to the well known torsional oscillations
detected at the surface. In what follows
we describe our results in detail. We include results on the
temporal variation as well as results
on the radial, latitudinal and time derivatives of the rotation rate.

\begin{figure} 
\plotone{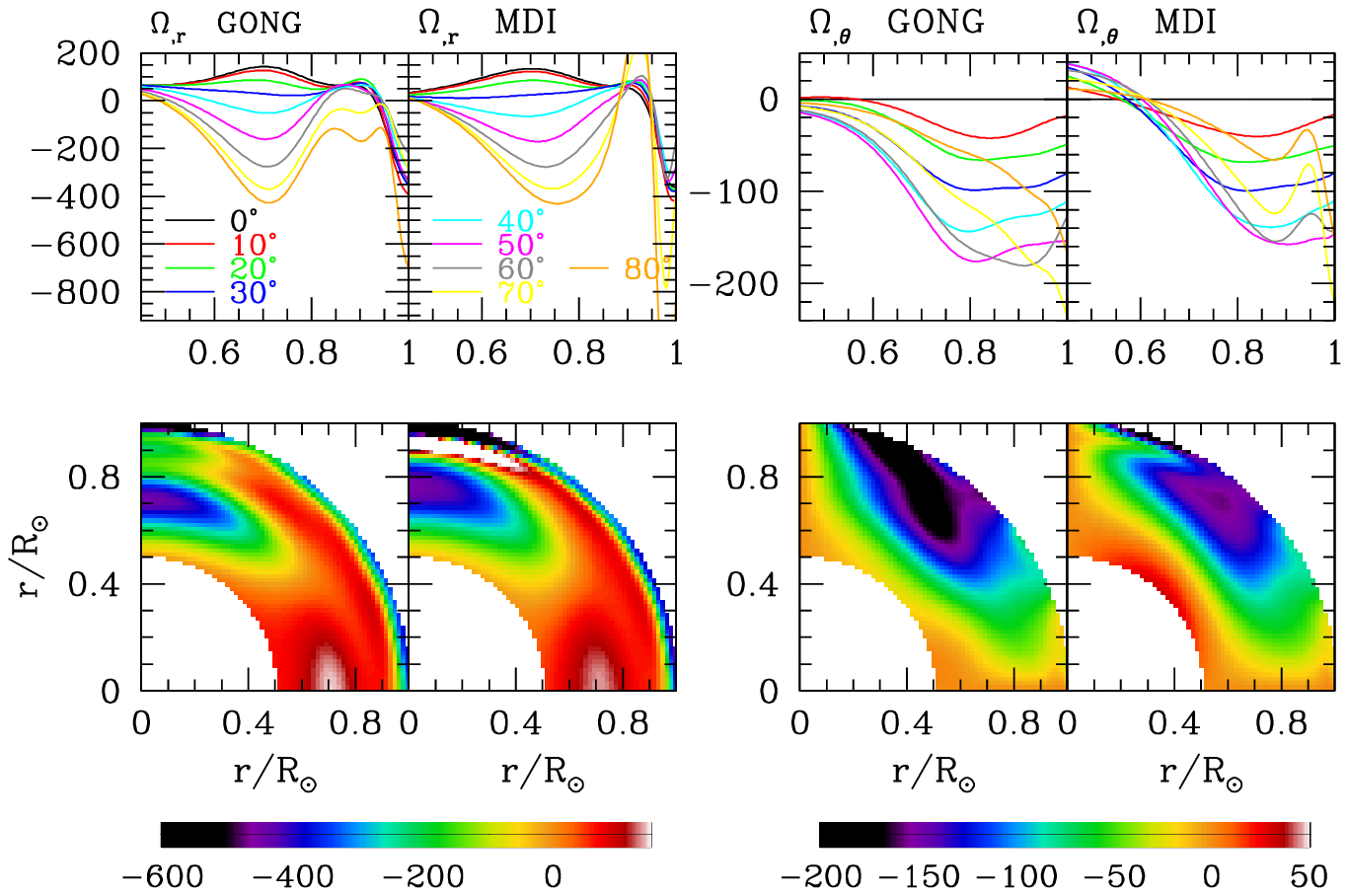}
\caption{The mean radial gradient, $\Omega_{,r}$, and latitudinal gradient,
$\Omega_{,\theta}$, of the solar rotation rate   obtained
using GONG and MDI data. Panels in the  upper row show the cuts at fixed latitude
as a function of radius, while those in the lower row shows the same results as
color-coded diagrams. All values are in units of nHz $R_\odot^{-1}$.}
\label{fig:shearav}

\end{figure}

Before discussing  temporal variations we examine the time-averaged
quantities. Since the mean rotation rate inferred by the inversion of
seismic data is well-known (e.g., Thompson et al.~1996; Schou et al.~1998),
we do not show the results here, but instead we only show the results for
the gradients, $\Omega_{,r}$ and $\Omega_{,\theta}$.
The upper panels of Fig.~\ref{fig:shearav}
show the cuts at fixed latitude, while the lower panels show the 
color-coded results as a function of both $r$ and $\theta$.

\begin{figure}
\plotone{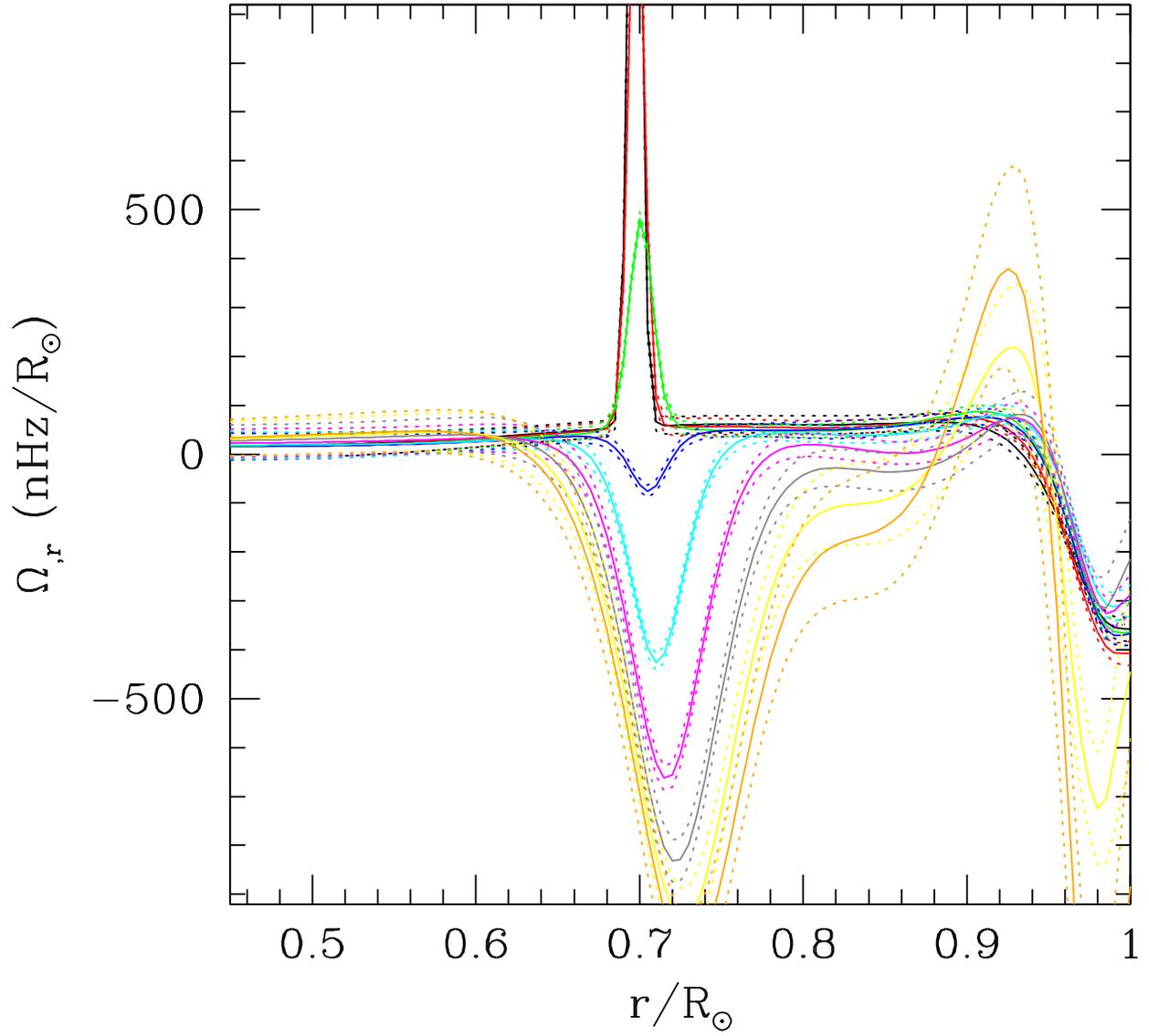}
\caption{The mean radial gradient, $\Omega_{,r}$, obtained  from
MDI data using the tachocline model of Antia et al.~(1998) shown at a
few selected latitudes. The color scheme is same as that in the upper
row of  Fig.~\ref{fig:shearav}. Errorbars are shown by dotted lines.}
\label{fig:sheartach}
\end{figure}

The radial gradient of the solar rotation rate
is large in the outer shear layer as well as in the tachocline, and 
is small in other regions. Below the outer shear layer, the radial
gradient is positive in the low latitude regions, and is negative in
all other regions wherever it can be determined reliably.
In the outer shear layer, we find an average negative  gradient of 
the order of a few 100 nHz $R_\odot^{-1}$ i.e, $10^{-15}$ m$^{-1}$ s$^{-1}$.
The value of $\Omega_{,r}$ is largely independent of latitude in the outer
shear layer, except perhaps  at very high latitudes.
It should be noted that the regularization used in carrying out
the  inversions smooths
out the radial variations near the tachocline region and hence the actual
radial gradient in the tachocline region is expected to be much larger.
If the tachocline model obtained by Antia et al.~(1998) is used to define
the rotation rate in the tachocline region, the gradient can be calculated
more precisely, and the results obtained using MDI data are shown in 
Fig.~\ref{fig:sheartach}.
It is clear that the gradient is much larger and more localized in the
tachocline region than is seen in Fig.~\ref{fig:shearav}.
 The exact shape of the curves, of course, depend on the
tachocline model used and may not have much significance.
The gradient in the latitudinal direction is significant only within the
convection zone, this is of course expected since this is where we
find differential rotation.  The  maximum value of the latitudinal
gradient  is comparable to the radial gradient
in the outer shear layer. This gradient is negative almost everywhere.
When studying the temporal evolution of the two gradients, we compare
the magnitude of the time-variation with the corresponding time-averaged values.

Although the pattern of variation of these gradients is similar for
results obtained with GONG and MDI data, the actual values show significant differences
in some regions. The results obtained from the two data sets agree reasonably well as
far as the
latitudinal gradient is concerned, though there are differences at high latitudes. 
There are
noticeable differences in the radial gradients obtained from the two data sets, particularly in the
outer shear layer.  Such differences between
results from GONG and MDI data  have been
seen even for the rotation rate itself (e.g., Schou
et al.~2002), and the reasons for this disagreement are not fully
understood. This issue is discussed further in connection with the
temporal variations of the radial gradient.

\subsection{The time-variation of the rotational velocity: the zonal flows}
\label{subsec:zonal}

Fig.~\ref{fig:cont8} shows the residual rotational velocity
$\delta v_\phi=\delta\Omega r\cos\theta$ obtained
using GONG data. Results are shown at a few representative radii in the convection zone.
The figure shows the expected, distinct,
bands of faster and slower-than-average rotation 
velocity that, at low latitudes,  move towards the equator with time.
At high latitudes, the bands seem to move towards the poles.
The slow band at high latitudes reaches its end near the poles around
2001, the time when the polar field reversed, while the fast band at high latitude, on the other hand,
appears to terminate around 1996 (and will  probably do so again in 2008) near the minimum of solar activity.
The transition between the equator-ward and pole-ward movements takes
place at a latitude of around $45^\circ$. 
At high latitudes it appears that the
12 years of GONG data have covered a little more than one period, but
at low latitudes the periodicity of the zonal-flow pattern is not obvious. 
The zonal-flow pattern in the outer shear layer appears to be more complex at low
latitudes than at high latitudes.
In this layer, the low-latitude  fast-moving bands from the two hemispheres
appear to have merged around 2000, but  seem to
split around 2005, and then merge again in 2007. 
A band of fast rotation appears near a latitude of $35^\circ$ in both
hemispheres around
2005 and we believe that these bands are a precursor to the next solar cycle.
In fact, a closer examination of the figure suggests that these bands have
a weak connection to the bands that started  at $50^\circ$ latitude around 1997
and then bifurcated around 2004 into one branch that moved  polewards and another
that moved towards the equator.
These bands should reach the equator and merge during the next
maximum phase of solar activity. Thus the band which started
near the $50^\circ$ latitude zone in 1997, i.e., just after the minimum phase of the activity, 
is expected to reach  
the equator during the maximum of the following cycle, and
will thus span a period of 15--18 years. The band of slow rotation
that started a few years after the last solar-activity maximum  around a latitude of
$50^\circ$  would reach the equator some time after the next minimum, again
spanning a period of about one and a half times the fiducial solar cycle.

Figure~\ref{fig:cont8} also shows the zonal flow-pattern obtained using GONG data
as cuts at a few representative latitudes. These can be used 
to determine how the zonal-flow pattern changes with radial distance.
At low latitudes there is a clear trend
of the pattern moving upwards with time and it is also clear that the
pattern penetrates through most of the convection zone, probably reaching
down to the base of the convection zone. At these low latitudes the average rate
of upward movement appears to be about $0.05R_\odot$ per year or about 1 m s$^{-1}$.
This follows from the fact that Fig.~\ref{fig:cont8}  shows the band moving from
close to the base of the convection zone to surface in about 6 years.
At a latitude of $60^\circ$ too the band of faster rotation
appears to penetrate nearly to the base of the convection zone near
the maximum of solar activity. 

\begin{figure}
\plotone{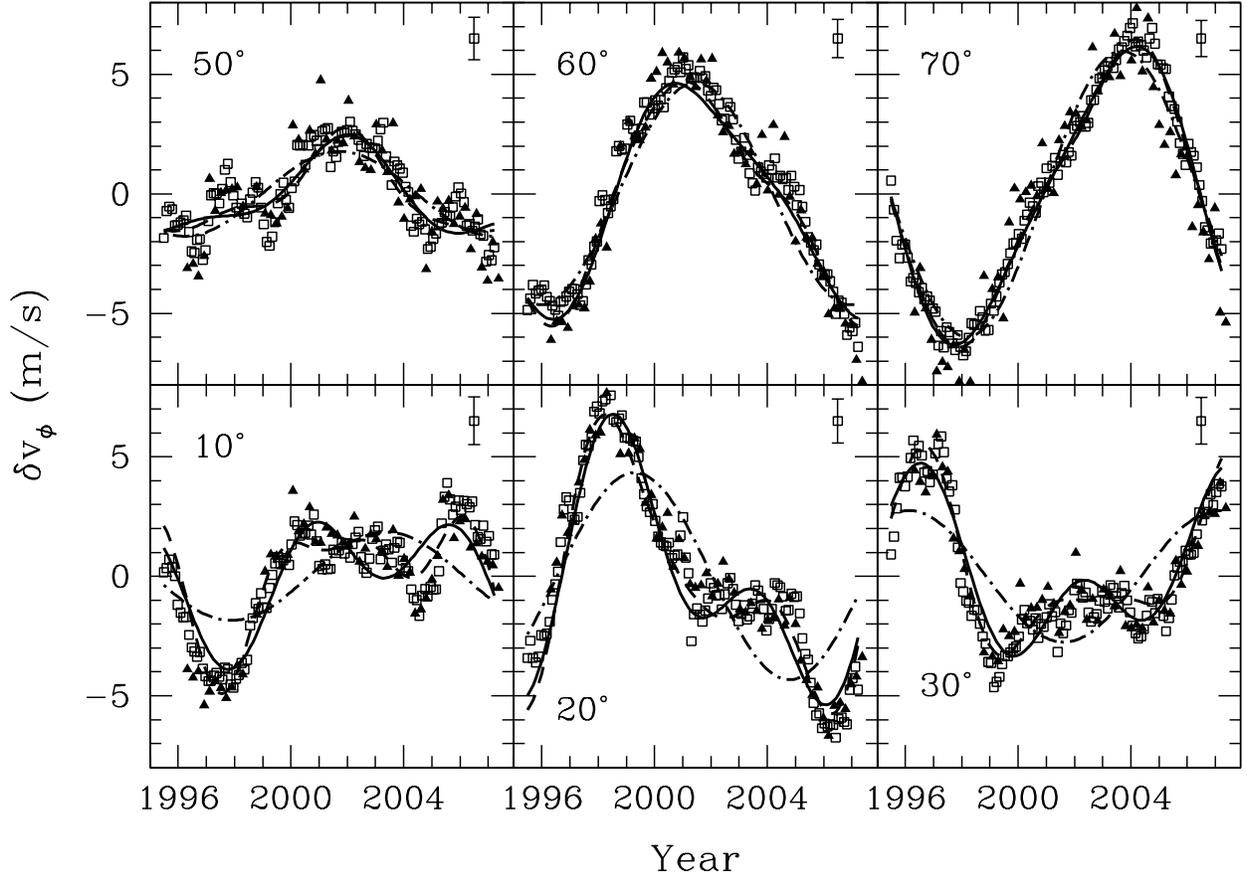}
\caption{The zonal-flow velocity as a function of time at different 
latitudes at $r=0.98R_\odot$. The latitudes are marked in each panel.
The squares show results obtained with GONG data, while triangles
show results with MDI data. The dash-dotted, solid and dashed lines show the fits
using  Eq.~(\ref{eq:harm}) 
to the GONG results with 1,2 and 3 terms respectively.
For clarity, errorbars are not shown on the points, however, a typical errorbar
is shown by a point in the upper right corner of each panel.
}
\label{fig:lat}
\end{figure}

The changing pattern of the zonal flows implies that the
maximum and minimum velocities of the flow occur at
different times  for different latitudes and depths.
This has also been found by  Schou (1999) and  Howe et al.~(2000a). 
Fig.~\ref{fig:lat} 
shows the zonal-flow velocity, $\delta v_\phi$, at different latitudes
as a function of time at $r=0.98R_\odot$. Both GONG and MDI results are 
shown and as can be seen, there is good agreement between the results obtained using data
from the two projects.
It is clear that the time-dependence of $\delta v_\phi$ changes with
latitude. While the high latitudes show a nearly sinusoidal variation
with a possible period of close to 11 years, the variation at low latitudes
is more complicated.
The Fourier transform of the results at all latitudes  show a peak at the lowest
frequency bin, which corresponds to a period of about 11 years.
The pattern at low latitudes, however, is clearly not sinusoidal. Thus if it is
periodic, higher harmonics must be present.
Vorontsov et al.~(2002) and Basu \& Antia (2003) had found the third harmonic
to be important. But those studies were based on data covering only about
half the solar cycle. With more data Howe et al.~(2005) and Basu \&
Antia (2006) found the second harmonic to be more significant than the third harmonic.
Clearly data from cycle 24 will be needed to establish the periodicity properly.
At a latitude of around $50^\circ$, the amplitude
of the flows is distinctly smaller than that at other latitudes.
This is close to the latitude where the switch from the equator-ward
movement to pole-ward movement of the flows occurs. Thus regions around this latitude
seem to separate the two branches of the zonal-flow pattern. 

\begin{figure} 
\plotone{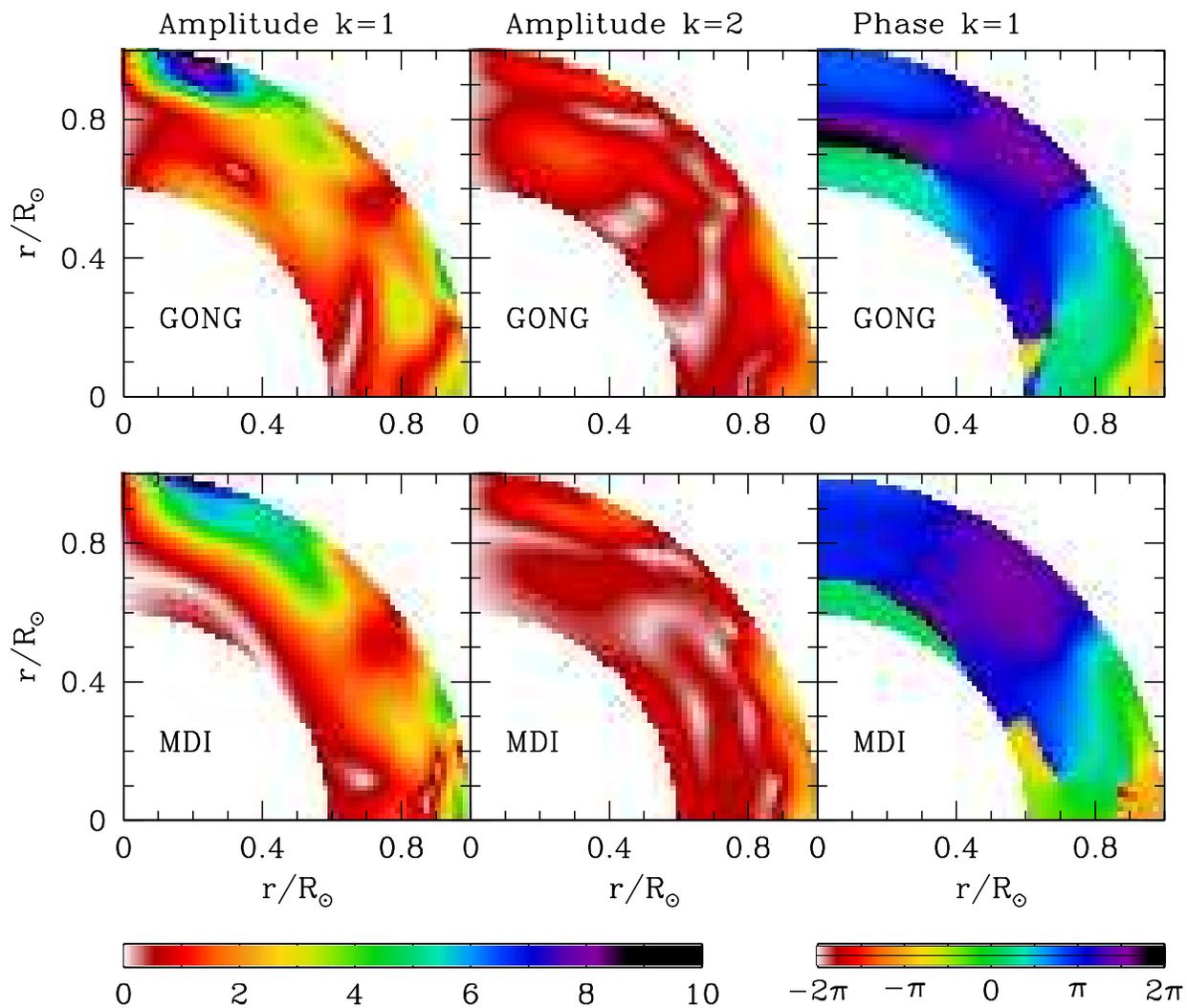}
\caption{Diagram showing amplitudes of the $k=1$ and $k=2$
(Eq.~\ref{eq:harm})
components and the phase of $k=1$ component of zonal-flow expansion as obtained using
GONG and MDI data.
For  panels that show the  amplitudes, the scale is in units of m s$^{-1}$.}
\label{fig:amp}
\end{figure}

While it is not possible to confirm the periodicity of the flows with current data sets,
it is tempting to assume a period
comparable to the average solar cycle. Thus we assume that the pattern is periodic with
a period of 11 years and fit it using Eq.~(\ref{eq:harm}) with
3 terms. The amplitudes of the first two terms are shown in Fig.~\ref{fig:amp}
for results obtained from both GONG and MDI data. Both these data sets give similar results
and it is clear that although the amplitude of the dominant $k=1$ component
decreases with depth, the pattern appears to penetrate to the base
of the convection zone. 
The highest amplitude, of up to 10 m s$^{-1}$, is seen in the outer $0.1R_\odot$  at high latitudes.
The amplitude is small in the  mid-latitude region around $45^\circ$.
It is clear that the $k=2$ harmonic is significant, however, its
magnitude is much smaller than that for the first term. The amplitude of the third term is found to
be even smaller and it is  difficult to estimate the amplitude reliably.
The relative magnitude of the various harmonics may depend on the origin
of the zonal flow pattern. 
In the very deep layers close to the tachocline, the errors
in the zonal-flow velocity estimates  are so large that it is difficult to
assign any significance to the fluctuating results obtained there. 
Below the convection zone the amplitude
of  both harmonics is very small, indicating that solar cycle-related
variations are not clearly seen in these layers.

Fig.~\ref{fig:amp} also shows the phase of the $k=1$ term
obtained from  both GONG and MDI data.
It is clear that at low
latitudes the phase changes rather steeply around $r=0.9R_\odot$.
At high latitudes while MDI results show very little phase variation with depth,
GONG results do show some variations. It is not clear if these differences are
significant. In the outer half of the convection zone, both GONG and MDI results 
appear to show a sharp transition around a latitude
of $40^\circ$.
This may mark the transition between the low and high latitude patterns.
There is good agreement between the GONG and MDI results.

We can extend the zonal flow pattern over longer time
intervals if we assume  that the fitted parameters do not change from cycle to cycle,
and the resulting pattern can be compared with predictions
from different dynamo models.  We shall be happy to make such figures available for
investigators who wish to make use of them.

\subsection{Temporal variations of the rotation-rate gradients}
\label{subsec:shear}

\begin{figure} 
\plotone{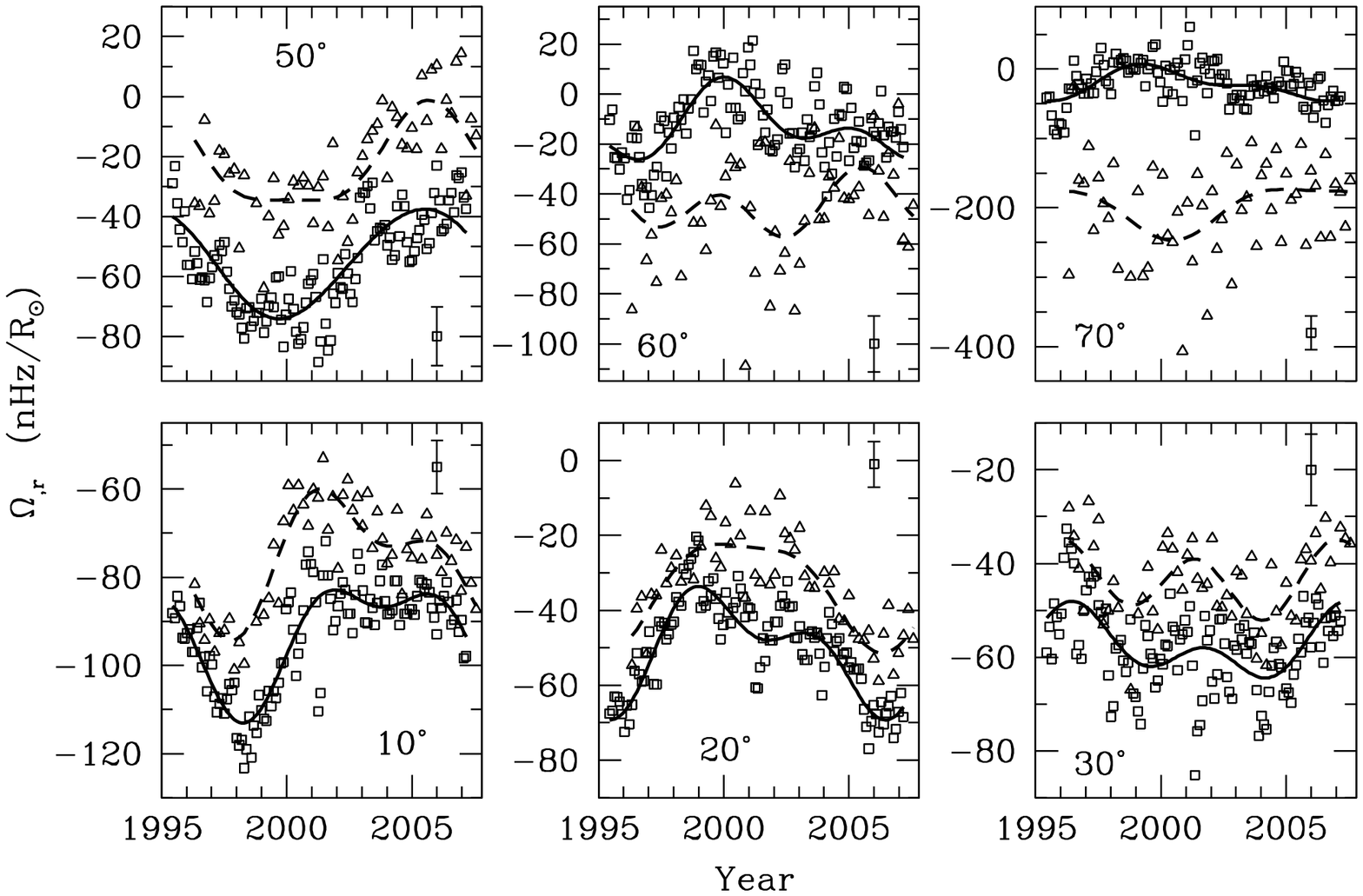}
\caption{The radial gradient, $\Omega_{,r}$
at a few selected latitudes is shown for $r=0.95R_\odot$.
The squares and triangles show results obtained using GONG and MDI data
respectively. The solid and dashed lines show fits obtained using equations
similar to Eq.~(\ref{eq:harm}) to these points.
For clarity, errorbars are not shown on the points, but a typical errorbar
 is shown by a point in a corner of each panel.}
\label{fig:sheardr}
\end{figure}

\begin{figure} 
\plotone{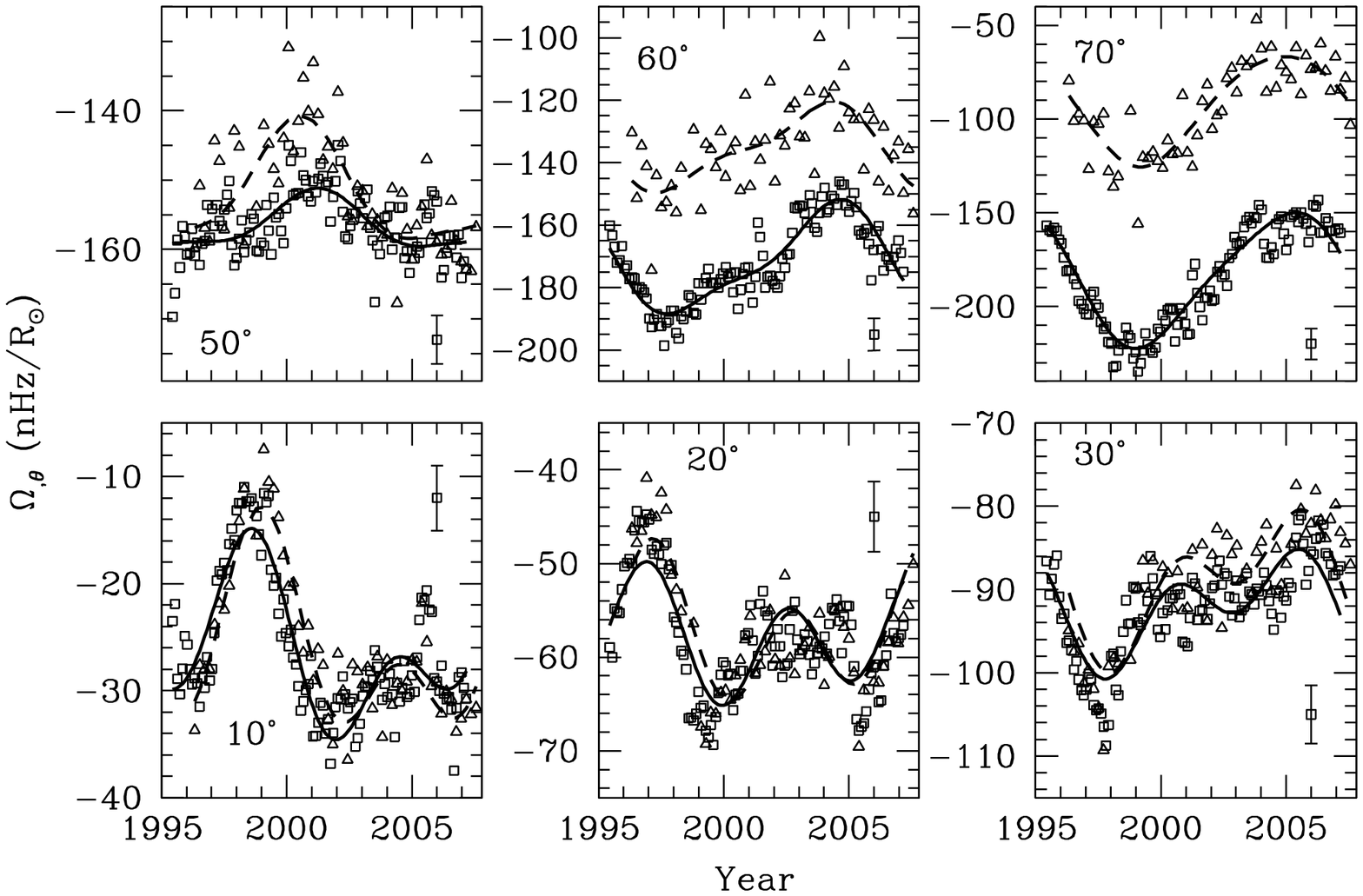}
\caption{The latitudinal gradient $\Omega_{,\theta}$
at a few selected latitudes is shown for $r=0.95R_\odot$.
The squares and triangles show the results obtained using GONG and MDI data
respectively. The solid and dashed lines show fits obtained using equations
similar to Eq.~(\ref{eq:harm}) to these points.
For clarity, errorbars are not shown on the points, but a typical errorbar
 is shown by a point in a corner of each panel.}
\label{fig:sheardth}
\end{figure}

We calculated the time variation of both radial and latitudinal
gradients by   numerical differentiation
of the  rotation-rate residuals that were obtained
using Eq.~(\ref{eq:zonal}).  
Figures~\ref{fig:sheardr} and \ref{fig:sheardth} show the two components of the
gradient  at different
latitudes as a function of time at $r=0.95R_\odot$.
There is a reasonable agreement between results obtained using GONG and MDI
data at this depth.
It is clear that there
is  significant time-variation in these gradients and  that the amplitude
of the variation is a sizable fraction of its average value.
At low latitudes, the temporal variation of the radial gradient turns
out to be about 20\% of its 
average value.
The relative variation in the latitudinal gradient is similar.
These changes are much larger than  the temporal variation in the rotation rate, which is
only about 0.1--1\% of its average value. As a result, we expect this
time-variation in the angular velocity gradients to 
play an important  role in workings of the solar dynamo. 

In order to establish a possible relation between the zonal-flow bands and
the magnetic cycle, we compare the pattern of temporal variations of
$\delta\Omega_{,r}$ and $\delta\Omega_{,\theta}$
at $r=0.95R_\odot$ with the butterfly diagram seen in the distribution
of sunspots. The results are shown in Fig.~\ref{fig:butter}.
We have chosen this depth because GONG and MDI results for the radial gradient do
not agree well in shallower layers.
We have multiplied the gradients by $\cos\theta$ in this and subsequent
figures in order to compensate for the increase in $\delta\Omega$
with latitude. This factor is already present in the residual rotation velocity $\delta v_\phi$
since it is part of the conversion from $\Omega$ to $v_\phi$.
As a result of this factor residuals in $v_\phi$ show little variation
in amplitude with latitude, and with
this factor, residuals in $\delta\Omega_{,r}$ or
$\delta\Omega_{,\theta}$ also show a similar behavior.
It can be seen from Fig.~\ref{fig:butter} that
the sunspots are mainly concentrated in regions where the radial gradient
is larger than its average value and where  the latitudinal gradient
is smaller than the average gradient.
Note, the mean values of both these gradients are negative at this depth,
and hence their magnitude will be larger in regions where $\delta\Omega_{,r}$ or
$\delta\Omega_{,\theta}$ are negative. Thus the sunspots are found to
occur in regions where the magnitude of radial gradient is smaller than
average and that of the latitudinal gradient is larger than average.
Clearly, the butterfly pattern is related to
zonal flows, as was, indeed, noted earlier by Snodgrass (1987).
Around 2005, when the equatorial zonal flow band of faster rotation splits,
the latitudinal gradient of rotation rate flips sign in low latitude
region.

\begin{figure} 
\epsscale{1.0}
\plotone{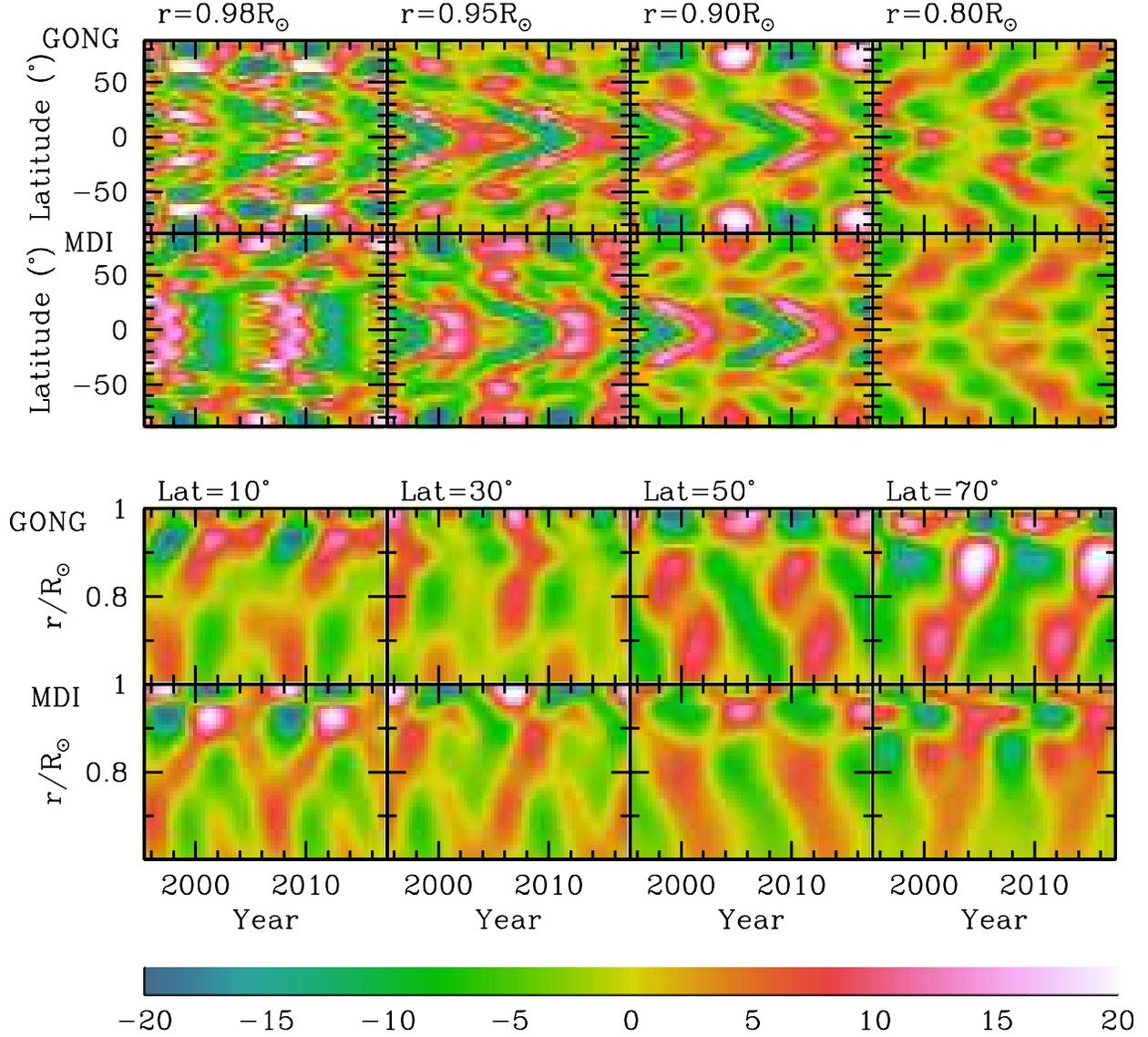}
\caption{Residuals in the radial gradient
obtained using GONG and MDI data at a few selected depths or latitudes. are shown.
We actually show  $\delta\Omega_{,r}\cos\theta$ to minimize latitudinal variation.
The scale is marked in units of nHz $R_\odot^{-1}$.}
\label{fig:domdrfit}
\end{figure}

\begin{figure} 
\plotone{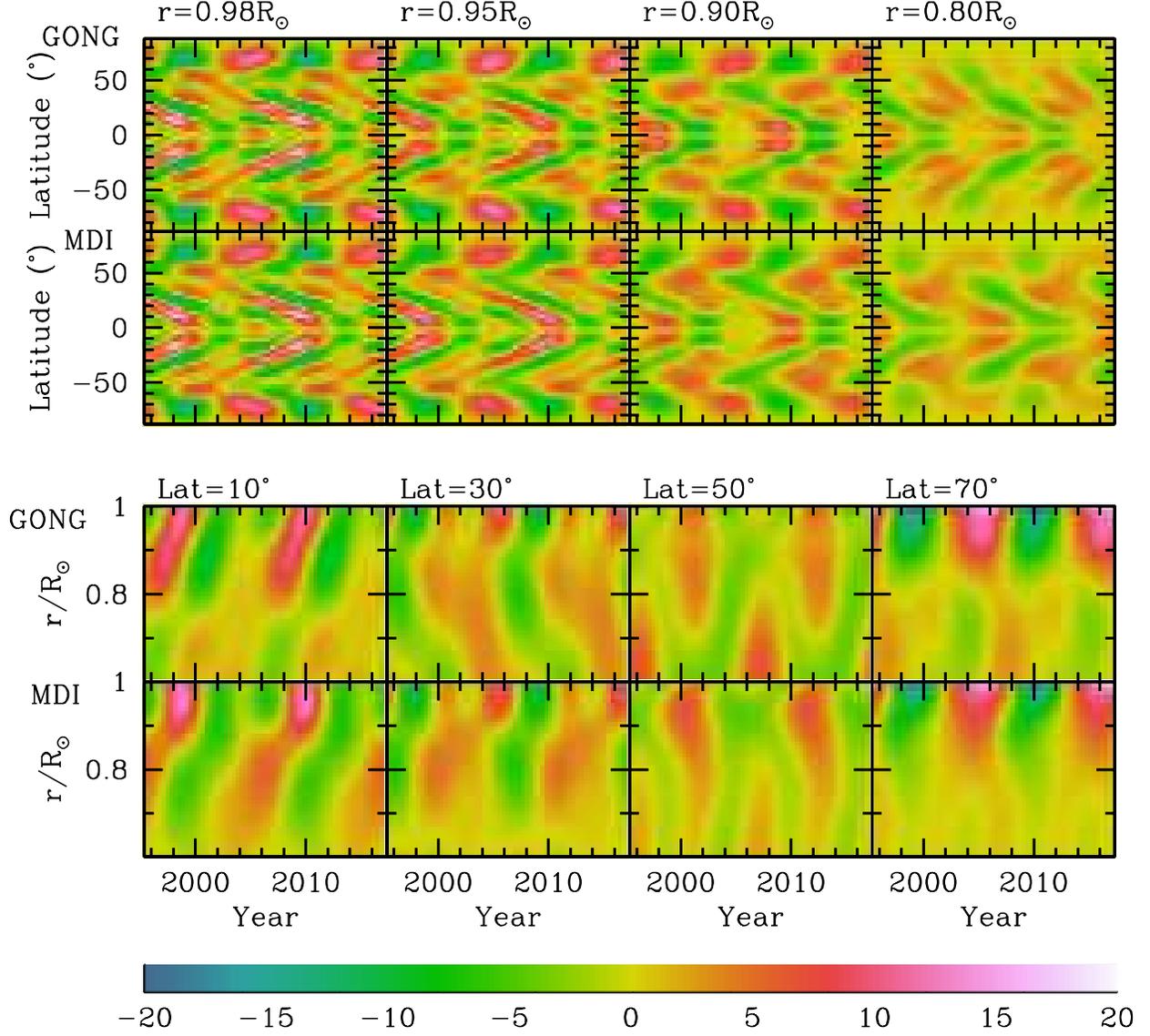}
\caption{Residual in the latitudinal gradient
obtained using GONG and MDI data at a few selected depths or latitudes.
We actually plot  $\delta\Omega_{,\theta}\cos\theta$
to minimize  latitudinal variation.
The scale is marked in units of nHz $R_\odot^{-1}$.}
\label{fig:domdthfit}
\end{figure}

The errors in our results on the  gradients are naturally expected to be
larger than those in
$\delta\Omega$. These large errors  give rise to some of the  fluctuations seen in Figs~\ref{fig:sheardr}
and \ref{fig:sheardth}.
As in the case of $\delta v_\phi$, these fluctuations can be smoothed  by fitting
 a periodic signal with a period of 11 years to the results  using Eq.~(\ref{eq:harm}).
The fitted results, extended to a period of 22 years, are shown in
Figs.~\ref{fig:domdrfit},\ref{fig:domdthfit}.
The pattern of bands of higher and lower than average gradients can be seen
in these figures too, with low latitude bands moving towards
the equator and the pattern once again changing at high latitudes.
In general there is reasonable agreement between GONG and MDI results, except
in the case of the radial gradient at $r=0.98R_\odot$ where there are significant differences
between the two results.
The agreement 
between GONG and MDI results is better in deeper layers, despite the fact that
the gradient itself is much smaller. As far as the latitudinal
gradient is concerned, the agreement between GONG and MDI results is good at all depths.
It has been known for some time that there are differences between GONG and MDI rotational-splitting data (Schou et al. 2002).
These differences result in the differences between results obtained with GONG and MDI data
about the outermost layers at all latitudes.
The most prominent difference is however, in the radial-gradient of the rotation rate at
high latitudes.
MDI data shows the presence of a jet like feature at high latitudes while GONG data do not
(e.g., Howe et al.~1998, Schou et al.~2002). 
Interestingly, the differences between the
GONG and MDI data do not affect the time variations of the
zonal-flow velocities and the results obtained from the two different
data sets agree reasonably well. This may indicate that the systematic
differences between the two data sets are largely independent of time.
Since we calculate the residual $\delta\Omega$ independently for GONG
and MDI the systematic errors cancel when the mean value is subtracted
in Eq.~(\ref{eq:zonal}).

\begin{figure} 
\epsscale{0.8}
\plotone{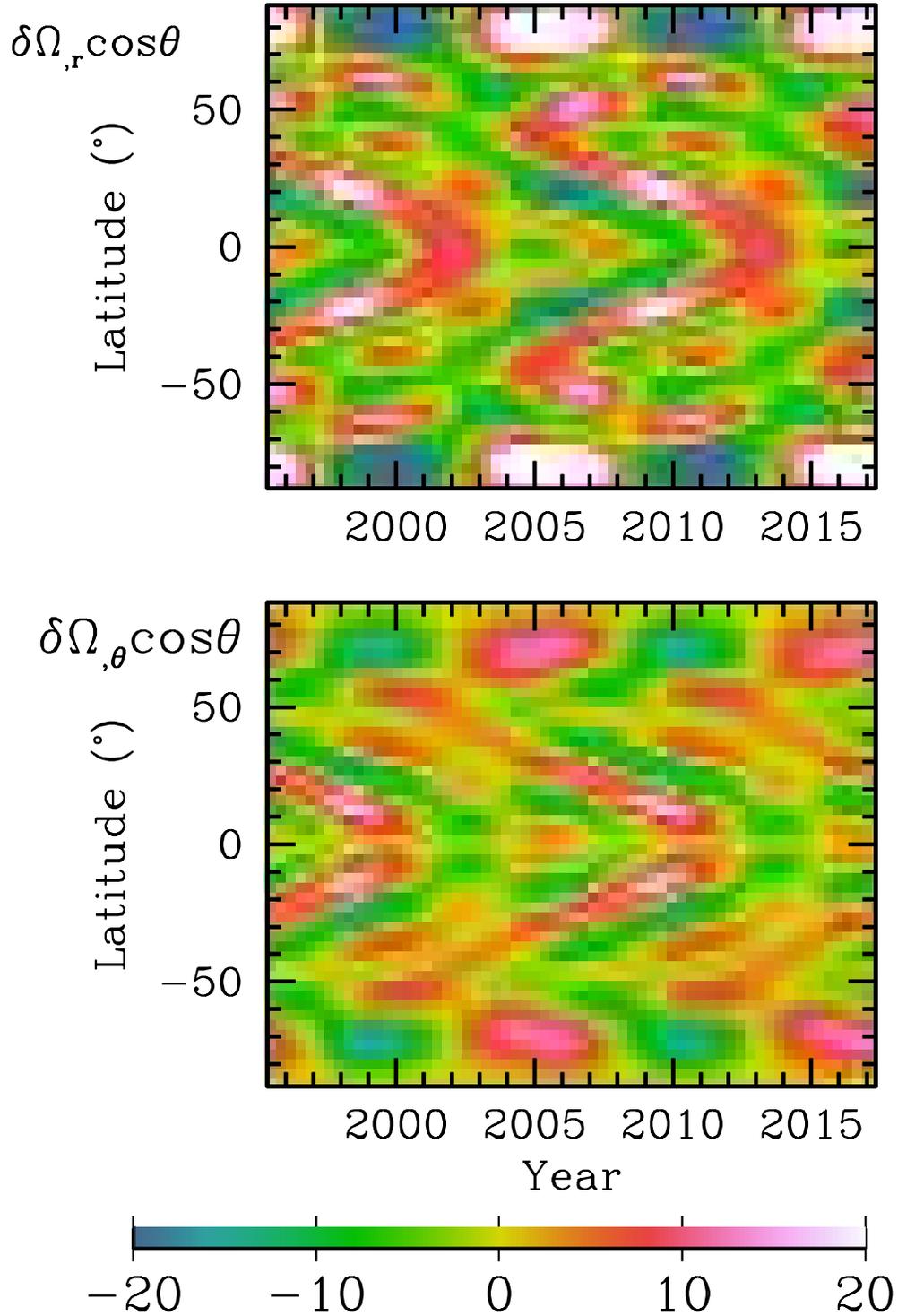}
\caption{Residual in the radial and latitudinal gradient
obtained at $r=0.98R_\odot$ using MDI data with $\ell<120$ modes.
We actually plot  $\delta\Omega_{,r}\cos\theta$ and
$\delta\Omega_{,\theta}\cos\theta$
to minimize  latitudinal variation.
The scale is marked in units of nHz $R_\odot^{-1}$.}
\label{fig:mdil120}
\end{figure}

It is tempting to speculate that the difference in radial gradient in outer
layers between the MDI and GONG results is because of the presence of higher
degree modes, in particular the f-modes, in MDI data sets. To test this, we 
repeated the calculations by removing these modes and find that the results remain
largely unaffected. Thus the difference between GONG and MDI results is unlikely
to be due to difference in resolution due to presence of extra modes and
possibly represents the difference in splitting coefficients themselves as
found by Schou et al.~(2002). While the zonal velocity, $\delta v_\phi$,
computed using the two data sets agree reasonably well,
the time variation of the
radial gradient of the inversions show significant differences. Thus the 
process of determining the radial gradient and its time variation appears
to  amplify the differences. In order to identify the modes that lead to this
difference, we did one more set of inversions for MDI data 
using only modes with $\ell<120$;  the resulting radial gradient is closer
to the result obtained using GONG data as can be seen from Fig.~\ref{fig:mdil120}. 
Thus it is clear that the differences are caused by the high-degree modes in the MDI sets.
Similar problems have been noticed in even-order splitting coefficients
from MDI (Antia et al.~2003). A closer look at Fig.~\ref{fig:domdrfit}
suggests that at $r=0.98R_\odot$ there was a rather sharp change around
1999 which is around the period when contact with SOHO was re-established.
This is similar to the results on asphericity found by Antia et al.~(2003)  and
results on solar radius variations found by Antia (2003). All these
could be due to some systematic error introduced during recovery
of SOHO. High-degree modes which are trapped in the outer layers
do not seriously affect the inversion
results in the deeper layers and hence, these layers  are not affected by the possible
problems with the MDI high-degree  modes. The re-analysis of MDI data has shown that
there are some removable problems with the currently available data sets (Larson \& Schou 2007).

The pattern at high latitudes is complicated and it is 
not possible to say with any certainty whether the pattern is moving towards the poles
or towards the equator. From Fig.~\ref{fig:domdrfit} it can be seen that
at $r=0.95R_\odot$, the band with lower than average radial gradient that started around a
latitude of $50^\circ$ in about 1995 ends at the equator around 2012, thus
spanning a period of about one and a half fiducial  solar cycles. This is  similar to the corresponding band in the
zonal-flow pattern. However, unlike the bands in the zonal-flow pattern,  these bands appear to
last for only about a solar cycle in the latitudinal-gradient pattern.
From the panels in Fig.~\ref{fig:domdrfit} that  show the radial dependence of
$\Omega_{,r}$ it also appears that
the pattern rises  upwards with time at low latitudes.
At the latitude of $30^\circ$
the amplitude of the variation in the radial gradient is rather small.
This is the latitude where the radial gradient in the lower convection
zone changes sign (cf., Fig.~\ref{fig:shearav}) and the magnitude of
temporally averaged $\Omega_{,r}$ is small too.
At high latitudes ($>45^\circ$) the bands showing the changes  in the
radial gradient penetrate to the base of the convection zone, while at
low latitudes the amplitude decreases below the outer
shear layer and there is a distinct phase shift around $r=0.9R_\odot$.
For the latitudinal gradient, the temporal variations
are generally significant only in the outer shear layer and  below a depth
of $0.1R_\odot$ this component is generally small (Fig.~\ref{fig:domdthfit}).

\subsection{The time derivative of $v_\phi$}
\label{subsec:acc}

\begin{figure} 
\epsscale{1.0}
\plotone{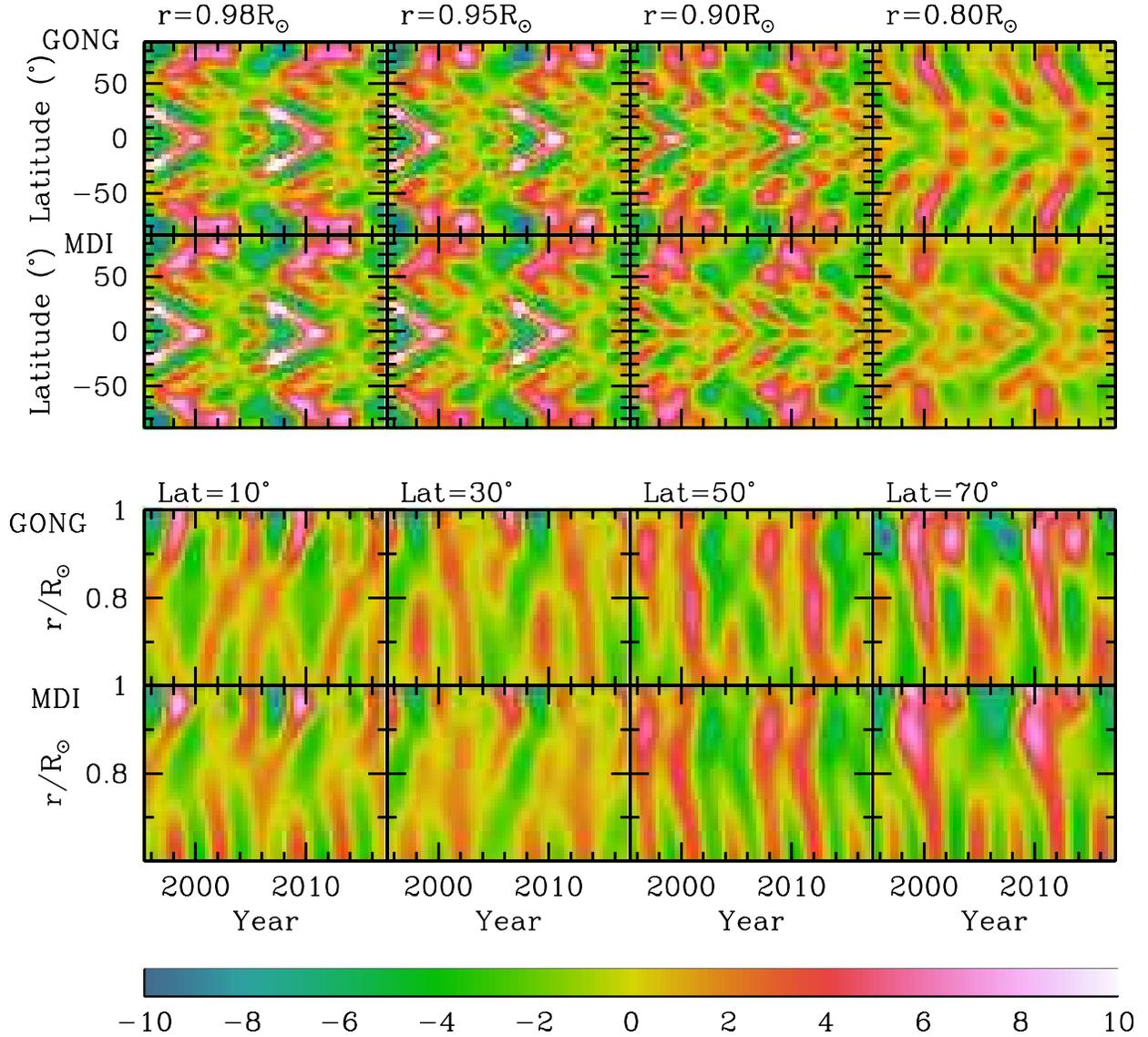}
\caption{The acceleration $(1/\omega_0)\partial v_\phi\partial t$
obtained using GONG and MDI data at a few selected depths or latitudes.
The scale is marked in units of m s$^{-1}$.}
\label{fig:dvrot}
\end{figure}

We evaluate the time-derivative of  $\delta v_\phi$, the zonal-flow velocity,
by differentiating the fits in Eq.~(\ref{eq:harm}). The results are
shown in Fig.~\ref{fig:dvrot}. As expected, the magnitude of the derivative
is  of the order of $\omega_0|\delta v_\phi|$, i.e.,  about $10^{-7}$ m s$^{-2}$.
Since this derivative is calculated by differentiating Eq.~(\ref{eq:harm}),
the pattern of the derivative is essentially similar to that of $\delta v_\phi$, except for a phase shift
of $\pi/2$. However, 
the higher harmonics make a  larger contribution
to the zonal-flow derivative than to the flow itself and as a result,
 there are  additional differences between
 the patterns that can be seen when  the pattern of the zonal-flow
derivative is compared to the pattern of the zonal flow. 
The  second and third harmonics appear to dominate 
 in the  deeper parts of the convection zone  near the tachocline, but
 since the signal
in this region is fairly weak even in $\delta v_\phi$, it is not clear if this
is significant.

\section{DISCUSSION AND CONCLUSIONS}
\label{sec:disc}

We have studied the time-variation of the solar-internal rotation rate and its gradients
over almost the entire period covering solar cycle 23 using
data from the GONG and MDI projects.  We used the 
rotation-rate residuals obtained by subtracting the time-averaged
rotation rate from that at each epoch at a given latitude and depth
to study the zonal-flow pattern and to determine
how it changes with time. These residuals have also been used to study the
radial and latitudinal gradients of the
zonal flows as well as the time-derivative of solar rotation velocity. Our main results are as follows:

\begin{enumerate}

\item The rotation-velocity residuals show the well known pattern similar to the observed pattern
of
torsional oscillations at the solar  surface. At low latitudes the results show  bands of faster and
slower-than-average rotation speed moving towards the equator as the solar
cycle progresses. At high latitudes on the other hand, these bands move towards the
poles. The transition between the equator-ward and pole-ward movement
occurs around a latitude of $45^\circ$. 

\item We find a precursor of the next solar cycle in the form of a band of fast 
rotation that appeared  at about the $35^\circ$ latitude near the surface around 2005 when
the Sun was approaching the activity minimum. 
We also find that the band
of  faster rotation found in the region near the equator appears 
to have bifurcated around the same time.

\item  The zonal flow pattern at low latitudes moves upwards, towards shallower depths, with an average speed of about 1 m/s.

\item The relative time variation of the radial gradient of the
rotation-rate is  about 20\%
of its average value, which is much larger than the relative variation in
the rotation rate itself. The time variation of the latitudinal gradient is large
only in the outer $0.1R_\odot$ of the Sun, the maximum change being about
20\% of the mean value.

\item The magnetic butterfly diagram coincides with band of larger than
average radial gradient and with smaller than average
latitudinal gradient at $r=0.95R_\odot$. Noting that both gradients are
negative, the sunspot activity bands occurs in regions of enhanced
latitudinal shear and diminished radial shear, as compared to the 
magnitude of the average shear.

\item The time derivative of the rotation velocity is about $10^{-7}$ m s$^{-2}$

\end{enumerate}

The equator-ward movement of the zonal-flow pattern at low latitudes is well known
(Schou 1999; Howe et al.~2000a; Antia \& Basu 2000; Vorontsov et al.~2002).
 The pole-ward migrating high latitude branch has also been noted earlier
 by Antia \& Basu (2001) and Ulrich (2001).
The transition between the equator-ward and pole-ward movements
occurs around a latitude of $45^\circ$. The differences between the low-latitude
and the high-latitude pattern also manifests in integrated quantities
like the rotational kinetic energy and angular momentum (Antia et al.~2008).
The time-variations of the rotational
kinetic energy integrated separately over low and high latitude regions also
differ.
The high-latitude regions  show a kinetic-energy variation that is correlated with
the solar activity throughout the convection zone. At the low latitudes,
on the other hand, the positive correlation between kinetic energy and solar activity
 exists only in the outer $0.1R_\odot$, while
the variations are anticorrelated with solar activity in the rest of the
convection zone.  Observations of magnetic
features at the solar surface also show 
an equator-ward movement at low latitudes (manifesting in the well-known
butterfly diagram) and a  pole-ward movement
at high latitudes (e.g., Leroy \& Noens 1983; Makarov \& Sivaraman 1989;
Erofeev \& Erofeeva 2000; Benevolenskaya et al.~2001).
Theoretical models of the solar dynamo, such as those of
Covas et al.~(2000, 2001), Bushby (2005) show these features, though they do differ in
detail, such as in the phase.  The
predicted butterfly diagrams of many models  (Jiang et al.~2007; Rempel 2006, etc.)
also show a low-latitude equator-ward moving branch and a high-latitude
pole-ward drifting branch.

\begin{figure} 
\plotone{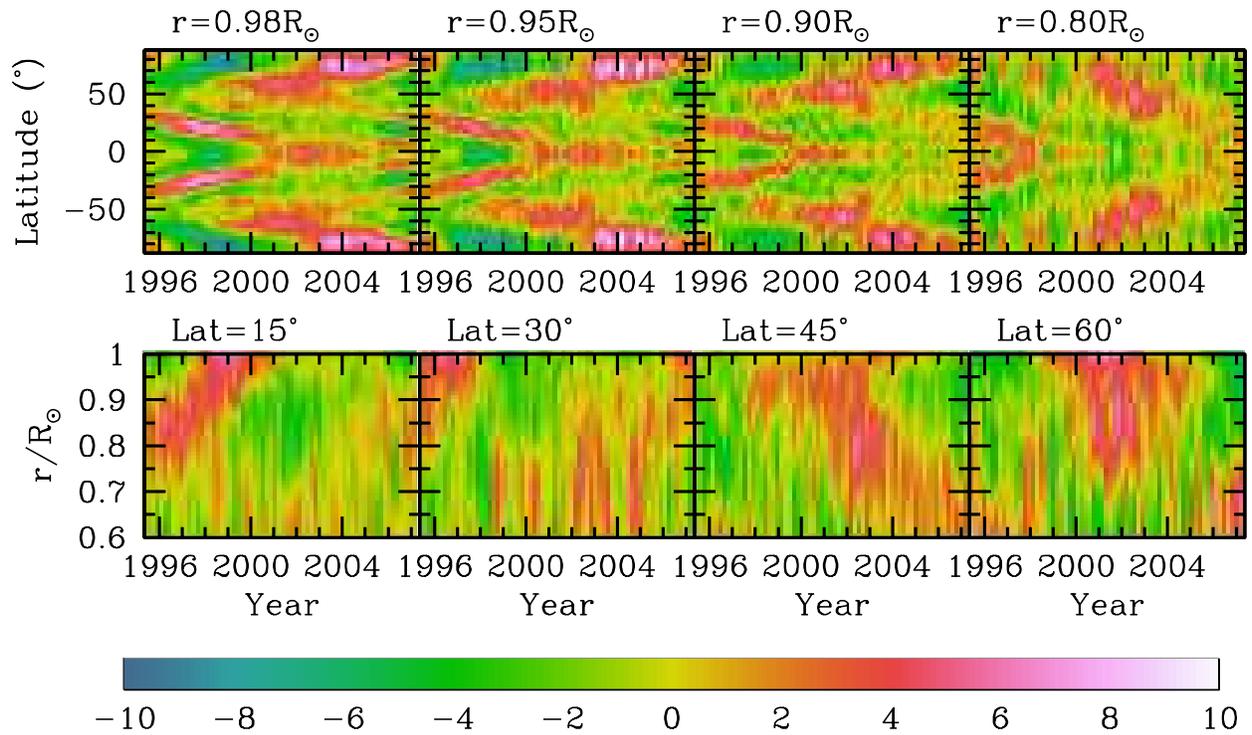}
\caption{Diagrams showing $\delta v_\phi$,
the residuals of the  rotation velocity obtained using
GONG data. We show the results 
at a few representative depths and latitudes as marked above the respective panels.
The scale is in m s$^{-1}$.}
\label{fig:cont8}
\end{figure}

The splitting of the  band
marking fast rotating region near the equator appears to be a new feature
and we are not aware of 
any other investigations that have noted this.
Although we have shown only GONG results, MDI data also show the splitting
of the bands.
The observations during the last solar minimum, however, does not show
this feature clearly. Although, a closer look at the Fig.~\ref{fig:cont8}
shows a faint signature of two bands merging at the equator around 1996,
albeit with a magnitude of $\delta v_\phi$ around 0.5 m s$^{1}$, which is
comparable to the error estimates and
about a factor of 5 lower than the corresponding value around 2006.
This difference could indeed be owing to cycle to cycle variation.
We clearly need more data to see the development of this pattern.
The near-surface behavior of zonal flows can be seen in the surface
rotation rate data from Mt.~Wilson (Ulrich 2001; Howe et al.~2006b)
that also covers the previous solar cycle. Since the errors in these
observations are larger, it is difficult to say whether the splitting of
the near equator band was seen in these results during the last cycle.

The rise of the zonal flow pattern with time has been seen in  sub-surface layers from
other measurements too.
 Conclusions similar to ours were drawn
by Komm, Howard \& Harvey~(1993) who compared the zonal flow pattern
obtained using the Doppler
measurements at the surface with that from magnetic features that
are believed to be anchored underneath the surface.
Our result that the zonal flows penetrate
through a good fraction of the convection zone, and  probably reach
the base of the convection zone appears to
contradict early inferences (Howe et al.~2000a; Antia \& Basu 2000)
that these flows penetrate only to a radius of $0.9R_\odot$.
The early results could have been due to the fact that the phase of the
zonal flow pattern at low latitudes changes around this depth.
With limited data this could have lead to the conclusion that the pattern does
not penetrate below this depth.
This phase change is seen in
variation of rotational kinetic energy in low latitudes too (Antia et al.~2008).
Our results are, however, similar to those of Vorontsov et al.~(2002), 
Basu \& Antia (2003) and Howe et al.~(2005).
Vorontsov et al.~(2002) also found that
a band of faster rotating elements appears to penetrate almost up to the
base of the convection zone at latitude of around $60^\circ$.
Dynamo models that assume that the tachocline is the seat of the
dynamo (e.g., Covas et al.~2000, 2001, 2004; Bushby 2005, etc.) also predict that the zonal
flow pattern should persist through the convection zone.

\begin{figure} 
\epsscale{0.8}
\plotone{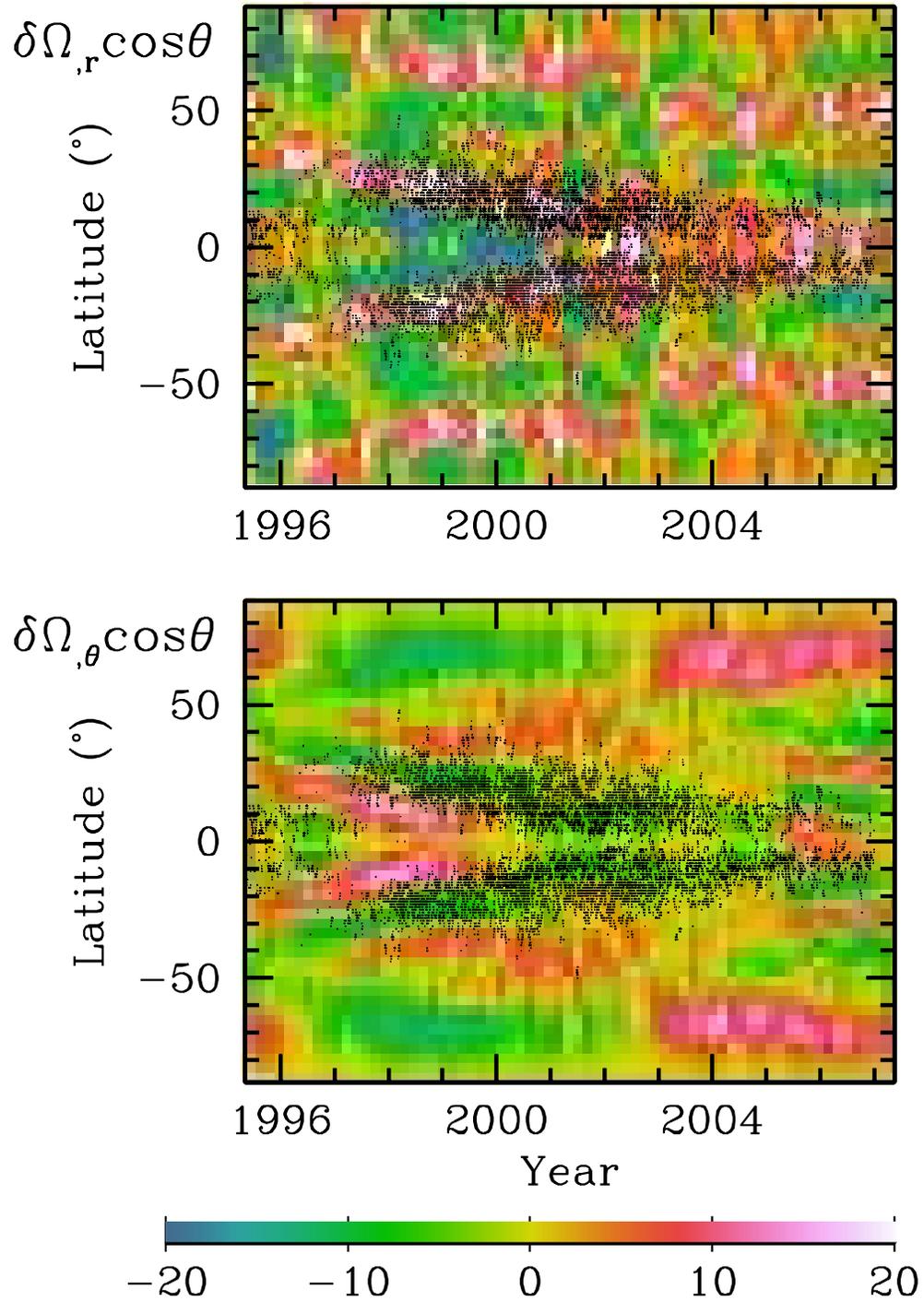}
\caption{The radial and latitudinal gradients, at $r=0.95R_\odot$
are superposed on the butterfly diagram showing the distribution
of sunspots which is obtained from the Greenwich sunspot data.
The scale is marked in nHz $R_\odot^{-1}$.}
\label{fig:butter}
\end{figure}

The possible link between time-dependent shear oscillation pattern and the
solar activity cycle was first emphasized by Snodgrass (1987). This becomes
evident from a striking resemblance of shear zones with the magnetic activity
bands revealed through the butterfly diagram (Fig.~\ref{fig:butter}).
Making the 
assumption that seat of the solar dynamo is likely to be located
within the shear layers, the observational
information that we have gained from the time-variation of radial
and latitudinal gradients of rotation rate should provide a valuable input
for understanding the dynamo mechanism.

We have hitherto been concerned largely with the time-variations of the global
rotation rate which is about 0.1--1\% of its mean value. The relative
temporal variations in the radial and latitudinal gradients of the rotation rate, however,
turn out to be larger,  20\% or more of their mean values. Clearly, the time-varying
shear pattern in the activity bands must play an important role in driving
the magnetic cycle. In fact, the time-derivatives of angular velocity
may be used either as valuable input for dynamo models or at any rate as
a constraint on them. Indeed, the time derivative of the differential rotation
has been effectively used in dynamo models such as those of Covas et al.~(2000)
and Rempel (2006).

It is illustrative to consider the azimuthal components of the induction and momentum
equations and neglect the presence of other velocity field such as
meridional circulation and dissipation, and write the equations
\begin{eqnarray}
{\partial B_\phi\over\partial t}&=& r\cos\theta{\bf B_p}.\nabla \Omega,
\\
\rho{\partial v_\phi\over\partial t}&=&{1\over 4\pi r\cos\theta}\nabla\cdot
\left({\bf B_p}(r\cos\theta B_\phi)\right),
\end{eqnarray}
where $\bf B_p$ and $B_\phi$ are respectively the poloidal and toroidal
components of the magnetic field, and $\rho$ is the density which is
assumed to be independent of time (e.g., Basu \& Antia 2000; Basu 2002). With the
available knowledge of the time-dependent radial and
latitudinal gradients of the angular velocity, it should be feasible to
deduce the temporal variation of the toroidal magnetic field for an
assumed configuration of the poloidal field. Equally, the force term
can also be estimated using the measured acceleration
$\partial v_\phi/\partial t$ provided we assume that the solar torsional
oscillations are driven mainly by the Lorentz force, although there will
be additional contributions from the temporally varying meridional
flow (e.g., Haber et al.~2002; Basu \& Antia 2003) which we neglect.
To an order of magnitude we can write
\be
{\partial v_\phi\over\partial t}\approx {|{\bf B_p}|B_\phi\over 4\pi r\rho}\;,
\label{eq:mag}
\ee
If the magnitude of the poloidal field is estimated through independent means,
it should be possible to infer the strength of the azimuthal field using
our knowledge of $\partial v_\phi/\partial t$. Thus, if the magnitude
of both poloidal and toroidal components are comparable, then the
magnetic field would be of the order of 1 G near the surface
increasing to 1000 G near the base of the convection zone.
The surface value is consistent with observations of the average
magnetic field at the surface (e.g., Ulrich
\& Boyden 2005).
Since the
magnetic field is generally concentrated in flux tubes with low filling factors, the
magnetic field in these flux tubes is expected to be much higher.
On the other hand, if the poloidal field is 100 times weaker than the toroidal field (e.g., Rempel
2007) then the toroidal field will be about 10 times the above estimate,
i.e., of the order of 10 kG near the base of the convection zone.

In conclusion, we emphasize that observations of the zonal flows of the
Sun now cover almost an entire solar cycle, and 
there are some features
of the flows seen at this solar minimum phase that had not been noticed in the earlier solar
minimum for which the helioseismic data were sparse. The further evolution of these features
should shed light on the differences between different solar cycles.
The current results are also  precise enough to provide
important inputs to or constraints for solar dynamo models. 

\acknowledgements

This work  utilizes data obtained by the Global Oscillation
Network Group (GONG) project, managed by the National Solar Observatory,
which is
operated by AURA, Inc. under a cooperative agreement with the
National Science Foundation. The data were acquired by instruments
operated by the Big Bear Solar Observatory, High Altitude Observatory,
Learmonth Solar Observatory, Udaipur Solar Observatory, Instituto de
Astrofisico de Canarias, and Cerro Tololo Inter-American Observatory.
This work also utilizes data from the Solar Oscillations
Investigation/ Michelson Doppler Imager (SOI/MDI) on the Solar
and Heliospheric Observatory (SOHO).  SOHO is a project of
international cooperation between ESA and NASA.
MDI is supported by NASA grants NAG5-8878 and NAG5-10483
to Stanford University. SB acknowledges partial support from
NSF grant ATM 0348837.
SMC thanks the Indian National Science Academy for support under the
INSA Honorary Scientist programme.

\clearpage

\clearpage

\clearpage

\clearpage

\clearpage

\clearpage

\clearpage

\clearpage

\clearpage

\clearpage

\clearpage

\clearpage

\end{document}